\documentclass[11pt,preprint]{aastex}

\usepackage{lscape}

\usepackage{epsfig,color}
\usepackage{amssymb,amsmath}
\usepackage{mathrsfs}
\usepackage{graphicx}
\usepackage{bm}


\def\oiii{[O~{\sc iii}]}
\def\feii{Fe~{\sc ii}}
\def\Hb{H$\beta$}

\def\dmfg1{$\mathscr{\dot M}_{f_{\rm g}=1}$}
\def\dmfg{$\mathscr{\dot M}_{f_{\rm g}}$}

\def\ledf1{$L_{\rm Edd}(f_{\rm g}=1)$}

\shorttitle{Measuring the Virial Factor in SDSS DR5 Quasars} \shortauthors{Liu et al.}

\begin{document}

\title{Measuring the Virial Factor in SDSS DR5 Quasars with Redshifted H$\beta$ and \feii\ Broad Emission Lines}

\author{H. T. Liu\altaffilmark{1,2,3}$^{\bigstar}$, Hai-Cheng Feng\altaffilmark{1,2,3,4}, Sha-Sha Li\altaffilmark{1,2,3}, and
J. M. Bai\altaffilmark{1,2,3}}

\altaffiltext{1} {Yunnan Observatories, Chinese Academy of Sciences, 396 Yangfangwang, Guandu District,
Kunming, 650216, People's Republic of China}

\altaffiltext{2} {Key Laboratory for the Structure and Evolution of Celestial Objects, Chinese Academy of Sciences, Kunming 650011, Yunnan, People's Republic of China}

\altaffiltext{3} {Center for Astronomical Mega-Science, Chinese Academy of Sciences, 20A Datun Road, Chaoyang District, Beijing, 100012, People's Republic of China}

\altaffiltext{4} {University of Chinese Academy of Sciences, Beijing 100049, People's Republic of China}

\altaffiltext{$^{\bigstar}$}{Corresponding author: H. T. Liu, htliu@ynao.ac.cn}

\begin{abstract}
  Under the hypothesis of gravitational redshift induced by the central supermassive black hole, and based on line widths and shifts of redward shifted \Hb\ and \feii\ broad emission lines for a sample of 1973 $z<0.8$ SDSS DR5 quasars, we measured the virial factor in determining supermassive black hole masses, usually estimated by the reverberation mapping (RM) method or the relevant secondary methods. The virial factor had been believed to be from the geometric effect of broad-line region. The measured virial factor of \feii\ is larger than that of \Hb\ for 98\% of these quasars. The virial factor is very different from object to object and for different emission lines. For most of these quasars, the virial factor of \Hb\ is larger than these averages that were usually used in determining the masses of black holes. There are three positive correlations among the measured virial factor of \Hb, dimensionless accretion rate and Fe {\sc ii}/\Hb\ line ratio. A positive three-dimensional correlation is found among these three quantities, and this correlation indicates that the virial factor is likely dominated by the dimensionless accretion rate and metallicity. A negative correlation is found between the redward shift of \Hb\ and the scaled size of broad-line region radius in units of the gravitational radius of black hole. This negative correlation will be expected naturally if the redward shift of \Hb\ is mainly from the gravity of black hole. Radiation pressure from accretion disk may be a significant contributor to the virial factor.

\end{abstract}


\section{INTRODUCTION}
Mass, $M_{\bullet}$, is an important fundamental parameter of black hole, and reliable measurement of $M_{\bullet}$
always will be a key issue of black hole related researches, such as the researches and understanding of the formation, growth, and evolution of the black holes in the Universe and the coevolution debates of supermassive black holes (SMBHs)
and host galaxies \citep[e.g.,][]{KH13}. Assuming virialized motion of clouds in broad-line region (BLR), the reverberation mapping (RM) method or the relevant secondary methods based on single-epoch spectra were widely used to measure $M_{\bullet}$ by a RM black hole mass $M_{\rm{RM}}=f v^2_{\rm{FWHM}} r_{\rm{BLR}}/G$ for active galactic nuclei (AGNs), where $f$ is the virial factor, $v_{\rm{FWHM}}$ is full width at half maximum of broad emission line, $r_{\rm{BLR}}$ is radius of BLR, and $G$ is the gravitational constant \citep[e.g.,][]{Pe04}. $f$ is commonly considered the main source of uncertainty in $M_{\rm{RM}}$. If the line width $v_{\rm{FWHM}}$ is replaced with $\sigma_{\rm{line}}$, the second moment
of emission line, $f$ becomes $f_{\sigma}$. The virial factor had been believed to be induced by the geometric effect of BLR. Based on the photoionization assumption \citep[e.g.,][]{BM82,Pe93}, $r_{\rm{BLR}}= \tau_{\rm{ob}} c/(1+z)$, where
$c$ is the speed of light, $z$ is redshift of source, and $\tau_{\rm{ob}}$ is the time lag observed between the broad-line and continuum variations. For non--RM AGNs with single-epoch spectra, $r_{\rm{BLR}}$ can be estimated with the radius--luminosity relation, i.e., the empirical $r_{\rm{BLR}}$--$L(\rm{5100\/\ \AA})$ relation for \Hb\ emission line, established on the basis of the RM AGNs, where $L(\rm{5100\/\ \AA})$ is AGN continuum luminosity at rest frame wavelength 5100 \AA\ \citep[e.g.,][]{Ka00,Be13,Du18b,Du19,Yu20}.

The RM observation researches have been made for more than 100 AGNs over the last several decades \citep[e.g.,][]{Ka99,Ka00,Ka07,Be06,Be10,Pe05,De10,Ba11,Ba15,Ha11,Po12,Du14,Du15,Du16,Du18a,Du18b,Pe14,Pe17,Wa14,Hu15,
Lu16,Xi18a,Xi18b,Zh19,Fe21a,Fe21b}. RM surveys have been run, such as the OzDES AGN spectroscopic RM project \citep{Ki15,Ho19} and the Sloan Digital Sky Survey (SDSS) spectroscopic RM project \citep{Sh15a,Sh15b,Sh16,Sh19,Gr17}.
The single-epoch spectrum was widely used to estimate $M_{\rm{RM}}$ for the SDSS quasars \citep[e.g.,][]{Hu08,Li19}
and the high-$z$ quasars \citep[e.g.,][]{Wi10,Wu15,Wa19}. However, the virial factor is very uncertain due to the
unclear kinematics and geometry of BLR \citep[e.g.,][]{Pe04,Wo15}. Radiation pressure of accretion disk has significant influences on the BLR clouds and the dynamics of clouds \citep[e.g.,][]{Ma08,Ne10,Kr11,Kr12,Na21}. The dynamics of
clouds can determine the three dimensional geometry of BLR \citep{Na21}. Thus, radiation pressure may be a contributor
to the virial factor. Radiation pressure was not considered in estimating $M_{\rm{RM}}$. Averages of $f \approx 1$
and/or $f_{\rm{\sigma}} \approx 5$ were derived based on the $M_{\bullet}-\sigma_{\ast}$ relation for the low-$z$
inactive and quiescent galaxies, where $\sigma_{\ast}$ is stellar velocity dispersion of galaxy bulge \citep[e.g.,][]{Tr02,On04,Pi15,Wo15}. $f \approx 1$ and/or $f_{\rm{\sigma}} \approx 5$ were usually used to estimate $M_{\rm{RM}}$ by the RM and/or the single-epoch spectra of AGNs. Thus, measuring $f$ and/or $f_{\rm{\sigma}}$
independently by a new method for individual AGNs is necessary and important to understand the physics of the BLR
and the issues related to black hole masses.

\citet{Li17} proposed a new method to measure $f$ based on the widths and shifts of redward shifted broad emission
lines for the RM AGNs. The Fe {\sc iii}$\lambda\lambda$ 2039--2113 UV line blend arises from an inner region
of the BLR \citep{MJ18}, and a lot of evidence shows that the UV blend originates close to the SMBH \citep[e.g.,][]{MJ21}.
Large values of $f$ are obtained from the widths and redward shifts of these UV blends, and an average of
$\langle f_{\rm{Fe III}}\rangle =14.3\pm 2.4$ is derived for 10 lensed quasars with a mean Eddington ratio of $\sim 0.8$ \citep{MJ20}. This average value is much larger than the widely accepted one of $f \approx 1$. However, the origin of redward shifts of broad emission lines are unclear, because the origin of broad emission lines in AGNs is not yet clear \citep[e.g.,][]{Wa17}. A tight correlation between broadening and redward shift of the Fe {\sc iii}$\lambda\lambda$ 2039--2113 blend for quasars in the BOSS survey supports the gravitational interpretation of its redward shift \citep{MJ18}. Alternative explanations, such as inflow, will need additional physics to explain the observed trend between broadening and redshift \citep{MJ18}. The redward shifts of the root mean square profiles of broad emission lines with respect to narrow emission lines and the BLR radii in Mrk 110 follow the gravitational redshift prediction \citep[see Fig. 3 in][]{Ko03}. The velocity-resolved time lags of H$\beta$ broad emission line for Mrk 50 and SBS 1518+593 show characteristic of Keplerian disk or virialized motion \citep{Ba11,Du18a}. A sign of the gravitational redshift $z_{\rm{g}}$ was found in a statistical sense for broad \Hb\ in the single-epoch spectra of SDSS DR 7 quasars \citep{Tr14}.

\citet{Hu08} suggested that inflow may generate the redward shifts of broad emission lines. Absorption lines that are redshifted with respect to the quasar's systemic velocity are an unambiguous signature of inflow \citep{Ru17}. Inflow generates the redward shifts of broad absorption lines relative to the quasar's systemic velocity determined from narrow emission lines, and the broad absorption and emission lines may be from different gas regions due to their distinct velocities \citep{ZS19}. Redward shifts of broad emission lines of H$\gamma$, H$\beta$, and He {\sc i} lines for Mrk 817 seemingly have an origin of outflow that is denoted by the redward asymmetric velocity-resolved lag profiles of these
lines \citep{Lu21}. However, the redward shifts of broad emission lines are commonly believed to be from inflow that will lead to the blueward asymmetric velocity-resolved lag profiles. This discrepancy implies that the redward shifts of broad emission lines do not originate from inflow. For each RM observation cycle of NGC 5548, \citet{Lu16} found that the variations of average 5100 $\rm{\AA}$ luminosity lead the changes of $r_{\rm{BLR}}$ by $\tau_{\rm{r-L}}=2.35$ yr, which is consistent with a dynamical timescale of $t_{\rm{BLR}}\approx 2.10$ yr for the BLR, and they obtained that the BLR could be jointly controlled by the radiation pressure of accretion disk and the central black hole gravity. \citet{Kr11} found that stable orbits of clouds in BLR exist for very sub-Keplerian rotation, for which the radiation pressure force contributes substantially to the force budget. Thus, the radiation pressure force might result in significant influence on the virial factor. In this work, SDSS DR5 quasars with redward shifted \Hb\ and \feii\ broad emission lines \citep[see Table 2 in][]{Hu08} are used to investigate the virial factor, relations between the virial factor and other physical quantities
for these quasars, and the origin of the redward shift of \Hb\ broad emission line.

The structure is as follows. Section 2 presents method. Section 3 describes sample selection. Section 4 presents discussion and conclusions. Throughout this paper, we assume a standard cosmology with $H_0=70 \rm{\/\ km \/\ s^{-1} \/\ Mpc^{-1}}$, $\Omega_{\rm{M}}$ = 0.3, and $\Omega_{\rm{\Lambda}}$= 0.7 \citep{Sp07}.



\section{METHOD}
A BLR cloud is subject to gravity of black hole, $F_{\rm{g}}$, and radiation pressure force, $F_{\rm{r}}$, due to central continuum radiation. Under the resultant force of $F_{\rm{t}}=F_{\rm{g}}+F_{\rm{r}}$, the total mechanical energy and angular momentum are conserved for the BLR clouds because $F_{\rm{g}}$ and $F_{\rm{r}}$ are central forces. Under various assumptions, $F_{\rm{r}}$ can be calculated for more than hundreds of thousands of lines, with detailed photoionization, radiative transfer, and energy balance calculations \citep[e.g.,][]{Da19}. In principle, $M_{\bullet}$ could be estimated by the BLR cloud motions as the numerical calculation methods give $F_{\rm{r}}$. However, the various assumptions may significantly influence the reliability of $F_{\rm{r}}$. Especially, many unknown physical parameters are likely various for different AGNs. Thus, a new method was proposed to measure $f$ when avoiding the numerical calculation
of $F_{\rm{r}}$ \citep{Li17}.

The virial factor formula in \citet{Li17} was derived from the Schwarzschild metric for a static cloud. In fact, the BLR clouds are not static, e.g., in the virialized motion. The gravitational redshift in the Schwarzschild space-time
for the BLR clouds can be expressed as \citep[see Equation 12 in][]{CB18}
\begin{equation}
   z_{\rm{g}} = \left(1-\frac{3G M_{\bullet}}{c^2 r_{\rm{BLR}}} \right)^{-1/2}-1,
\end{equation}
where the gravitational and transverse Doppler shifts are taken into account. $M_{\bullet}$ is estimated as
\begin{equation}
  M_{\bullet}=\frac{1}{3}G^{-1}c^2 r_{\rm{BLR}} \left[ 1-\left( 1+z_{\rm{g}}\right)^{-2} \right],
\end{equation}
and the first order approximation is
\begin{equation}
  M_{\bullet}= \frac{2}{3} G^{-1} c^2 z_{\rm{g}} r_{\rm{BLR}},
\end{equation}
as $z_{\rm{g}} \ll 1$ or $r_{\rm{g}}/r_{\rm{BLR}}\ll 1$ for optical broad emission lines (the gravitational
radius $r_{\rm{g}}= GM_{\bullet}/c^2$).

Here, Equation (3) is the same as Equation (3) of \citet{MJ18}, in which the weak field limit of the Schwarzschild metric was assumed. At the optical BLR scales, the Schwarzschild metric is valid and matches the weak field limit. Equation (1) is valid for a disklike BLR \citep[see][]{CB18}. The disklike BLR is preferred by some RM observations of AGNs, e.g., NGC 3516 \citep[e.g.,][]{De10,Fe21a}, and the VLTI instrument GRAVITY observations of quasar 3C 273 \citep{St18}. For rapidly rotating BLR clouds, the relativistic beaming effect can give rise to a profile asymmetry with an enhanced blue side in broad emission lines, i.e., blueshifts of broad emission lines \citep{Me89}. Thus, the relativistic beaming effect should be neglected for the redward shifted broad emission lines, which should be dominated by the gravitational redshift and transverse Doppler effects. The factor of 2/3 in Equation (3) results from correcting the transverse Doppler shift, which is essentially that the moving clock becomes slower. The factor of 2/3 does not appear in the formulas used to estimate $M_{\bullet}$ in \citet{Ko03} and \citet{Li17}, because they did not consider the transverse Doppler shift. If $M_{\bullet}$ estimated with Equation (2) is equal to $M_{\rm{RM}}$, we have the virial factor
\begin{equation}
   f = \frac{1}{3} \frac{c^2}{v^2_{\rm{FWHM}}} \left[1-\left(1+z_{\rm{g}} \right)^{-2} \right].
\end{equation}
If $v_{\rm{FWHM}}$ is replaced with $\sigma_{\rm{line}}$, $f$ becomes $f_{\sigma}$.

Because $r_{\rm{BLR}}\gg r_{\rm{g}}$ for optical broad emission lines, Equation (1) can give the multi-broad-line approach of measuring $M_{\bullet}$ as
\begin{equation}
  M_{\bullet}\cong \frac{2}{3}G^{-1} c^2 \Delta z_{\rm{i,j}} \left(\frac{1}{r_{\rm{BLR,i}}}
  - \frac{1}{r_{\rm{BLR,j}}}\right)^{-1},
\end{equation}
where $\Delta z_{\rm{i,j}}=z_{\rm{i}}-z_{\rm{j}} = z_{\rm{g,i}}-z_{\rm{g,j}}$ is the redshift difference between the
broad lines $i$ and $j$ with the relevant BLR radius $r_{\rm{BLR,i}}$ and $r_{\rm{BLR,j}}$. Here, Equations (2), (3),
(4), and (5) have the factor of 2/3 more than Equations (4), (5), (9), and (7) derived in \citet{Li17}, respectively.
The reliability of the redward shift method was confirmed by the consistent masses estimated from their Equations (4) and (7) based on 4 broad emission lines for Mrk 110 \citep{Li17}. Thus, Equations (2) and (5) in this work can also
give consistent black hole masses for Mrk 110. The RM observations of multi broad emission lines for AGNs might further
test the reliability of this method, based on Equations (2) and (5). Hereafter, $M_{\rm{RM}}$ denotes $M_{\bullet}$ measured with the RM method and/or the relevant secondary methods, and $f_{\rm{g}}$ denotes the virial factor that comes from the geometric effect of BLR.

\section{SAMPLE SELECTION}
\citet{Hu08} reported a systematical investigation of optical \feii\ emission in a large sample of 4037 $z < 0.8$ quasars selected from the SDSS DR5, for which they had developed and tested a detailed line-fitting technique, taking into account the complex continuum and narrow and broad emission line spectra. The line widths and redward velocity shifts of the \feii\ and \Hb\ spectra are given in Table 2 of \citet{Hu08}. On the basis of $\Delta v - \sigma(\Delta v)> 0$, where $\Delta v$
is the redward velocity shift for the \Hb\ and \feii\ broad emission lines (i.e., $\Delta v>0$), and $\sigma(\Delta v)$ is the error of $\Delta v$, 1973 quasars are selected out of these 4037 quasars as our sample. This selection condition makes sure that the velocity shift is larger than zero within $1\sigma$ uncertainties. If $\Delta v - \sigma(\Delta v)\leq 0$,
it is possible that the velocity shift is redshift, blueshift, or no-shift. Thus, the redward velocity shift seems much
less reliable if $\Delta v - \sigma(\Delta v)\leq 0$, and this selection condition of $\Delta v - \sigma(\Delta v)> 0$
seems reasonable.

Because the empirical $r_{\rm{BLR}}$--$L(\rm{5100 \/\ \AA})$ relation is established for broad emission line \Hb, the relevant researches of the virial factor are made mainly with the broad \Hb\ line. Some physical quantities are taken
from Table 2 in \citet{Hu08}, including the cosmological redshift of source: $z$, $v_{\rm{FWHM}}$(\Hb), $\sigma_{\rm{line}}$(\Hb), the redward velocity shift of broad \Hb: $\Delta v$(\Hb), $v_{\rm{FWHM}}$(\feii), $\Delta v$(\feii), $L(\rm{5100 \/\ \AA})$, the black hole mass, the Eddington ratio, and the line ratio of Fe {\sc ii} to \Hb. The bolometric luminosity in the Eddington ratio was estimated in \citet{Hu08} using $L_{\rm{bol}}=9L(\rm{5100 \/\ \AA})$ \citep{Ka00}. The details of sample are listed in Table 1. The virial factor is estimated by Equation (4), and the
relevant values for the \Hb\ and \feii\ broad emission lines are listed in Table 1. The dimensionless accretion rate $\mathscr{\dot M}_{f_{\rm{g}}} = L_{\rm{bol}}/L_{\rm{Edd}}(f_{\rm{g}})/\eta$, where $\eta$ is the efficiency of converting rest-mass energy to radiation, $L_{\rm{bol}}$ is the bolometric luminosity, $L_{\rm{Edd}}$ is the Eddington luminosity, $f_{\rm{g}}=1$ for $v_{\rm{FWHM}}$, and $f_{\rm{g}}=5.5$ for $\sigma_{\rm{line}}$. Here, we adopt $\eta =0.038$ \citep{Du15}.

\begin{landscape}
\begin{deluxetable}{cccccccccccccccc}
  \tablecolumns{16}
  \setlength{\tabcolsep}{1.6pt}
  \tablewidth{0pc}
  \tablecaption{The relevant parameters for 1973 quasars in SDSS DR5}
  \tabletypesize{\scriptsize}
  \tablehead{\colhead{Designation} & \colhead{$z$} & \colhead{$\frac{v_{\rm{FWHM}}(\rm{H\beta})}{\rm{km \/\ s^{-1}}}$}  & \colhead{$\frac{\sigma_{\rm{H\beta}}}{\rm{km \/\ s^{-1}}}$} & \colhead{$\frac{\Delta v(\rm{H\beta})}{\rm{km \/\ s^{-1}}}$} & \colhead{$\frac{v_{\rm{FWHM}}(\rm{Fe II})}{\rm{km \/\ s^{-1}}}$} & \colhead{$\frac{\Delta v(\rm{Fe II)}}{\rm{km \/\ s^{-1}}}$}& \colhead{$\log L$} & \colhead{$\frac{M_{\rm{RM}}}{10^7 M_{\odot}}$} & \colhead{$\frac{L_{\rm{bol}}}{L_{\rm{Edd}}}$} & \colhead{$R_{\rm{Fe}}$}  & \colhead{$f$(\Hb)} & \colhead{$f_{\sigma}$(\Hb)} & \colhead{$f$(\rm{\feii})} & \colhead{$\log \mathscr{\dot M}_{f_{\rm{g}}}$}
  & \colhead{$\frac{r_{\rm{BLR}}}{r_{\rm{g}}}$} \\
  \colhead{(1)}  &\colhead{(2)}  &\colhead{(3)}  &\colhead{(4)}  &\colhead{(5)}  &\colhead{(6)}  &
  \colhead{(7)}  &\colhead{(8)}  &\colhead{(9)}  &\colhead{(10)} &\colhead{(11)} & \colhead{(12)} &\colhead{(13)}&\colhead{(14)}&\colhead{(15)}&\colhead{(16)} }

  \startdata
 000011.96+000225.3 & 0.4784 & 3135.8$\pm$62.6 & 1893.5 & 542.3$\pm$35.4 & 1898.6$\pm$74.1 & 387.6$\pm$38.4 & 44.74 & 28.1 & 0.140 & 1.318 & 11.0$\pm$0.8 & 30.2$\pm$3.7 & 21.5$\pm$2.7 & 0.57 & 4535.0 \\
 000111.19 - 002011.5 & 0.5173 & 3666.5$\pm$125.3 & 2267.1 & 486.9 $\pm$66.3 & 2496.5$\pm$406.2 & 1082.1$\pm$156.9 & 44.60 & 32.2 & 0.088 & 0.609 & 7.2$\pm$1.1 & 18.9$\pm$3.2 & 34.5$\pm$15.9 & 0.37 & 3159.8  \\
 000131.42+144610.6 & 0.5309 & 5054.5$\pm$306.4 & 2299.3 & 481.9$\pm$103.9 & 3359.2$\pm$658.1 & 1595.2$\pm$214.5 & 44.44 & 25.8 & 0.077 & 0.640 & 3.8$\pm$0.9 & 18.2$\pm$4.4 & 28.0$\pm$17.1 & 0.31 & 3069.8 \\
... & ... & ... & ... &  ... & ... &  ... & ... & ... & ... & ... & ... & ... & ... & ... & ...\\

\enddata
\tablecomments{Column 1: object name; Column 2: redshift; Column 3: $v_{\rm{FWHM}}$ of \Hb\ broad emission line; Column 4: $\sigma_{\rm{line}}$ of \Hb\ broad emission line; Column 5: the redward velocity shift of \Hb; Column 6: $v_{\rm{FWHM}}$ of \feii\ broad emission line; Column 7: the redward velocity shift of \feii; Column 8: logarithm of $L(\rm{5100 \/\ \AA})$ in units of $\rm{erg \/\ s^{-1}}$; Column 9: the black hole mass; Column 10: the Eddington ratio; Column 11: the Fe {\sc ii}/\Hb\ line ratio; Column 12: the virial factor estimated from $v_{\rm{FWHM}}$ of \Hb; Column 13: the virial factor estimated from $\sigma_{\rm{line}}$ of \Hb; Column 14: the virial factor estimated from $v_{\rm{FWHM}}$ of \feii; Column 15: logarithm of $\mathscr{\dot M}_{f_{\rm{g}}=5.5}$; Column 16: $r_{\rm{BLR}}$ in units of $r_{\rm{g}}$, where $r_{\rm{BLR}}=22.3L^{0.69}_{44}$ light-days with $L_{44}=L(\rm{5100 \/\ \AA})/(10^{44}\/\ \rm{erg \/\ s^{-1}})$. Columns 2--11 are taken from Table 2 of \citet{Hu08} or converted from the relevant quantities in Table 2 of \citet{Hu08}. \\(This table is available in its entirety in machine-readable form.)}
\end{deluxetable}
\end{landscape}

\section{ANALYSIS AND RESULTS}
The Spearman's rank correlation test shows that the virial factor is positively correlated with the dimensionless
accretion rate $\mathscr{\dot M}_{f_{\rm{g}}=5.5}$ for these 1973 quasars (see Figure 1 and Table 2). The virial factor
and $\mathscr{\dot M}_{f_{\rm{g}}=5.5}$ are related to the line width, and this line width dependency may result in a
false correlation between them. The partial correlation analysis gives a confidence level of 99.99\% for the positive correlation of $\log f_{\sigma}$--$\log \mathscr{\dot M}_{f_{\rm{g}}=5.5}$ when excluding the dependence on the line
width $\sigma_{\rm{line}}$. Since the virial factor may be affected by $F_{\rm{r}}$, it is possible that the virial
factor is correlated with $L(\rm{5100 \/\ \AA})$. So, we analyze the virial factor and $L(\rm{5100 \/\ \AA})$, and find
no correlation between them (see Figure 2). Thus, the positive correlation exists between the virial factor and $\mathscr{\dot M}_{f_{\rm{g}}=5.5}$. This positive correlation is qualitatively consistent with the logical expectation
when the overall effect of $F_{\rm{r}}$ on the BLR clouds is taken into account to estimate $M_{\rm{RM}}$. In addition, $f_{\rm{\sigma}}>f_{\rm{g}}=5.5$ and $f>f_{\rm{g}}=1$ for \Hb\ in most of quasars (see Figure 1).

In order to test the gravitational origin of the redward velocity shift of broad emission line, we compare
$\Delta v$(\Hb) to $r_{\rm{BLR}}/r_{\rm{g}}(f_{\rm{g}}=5.5)$, the BLR radius in units of the gravitational
radius of black hole. The Spearman's rank correlation test shows negative correlation between the velocity shift and $r_{\rm{BLR}}/r_{\rm{g}}(f_{\rm{g}}=5.5)$ (see Figure 3$a$ and Table 2). This negative correlation is qualitatively consistent with the expectation when $\Delta v$(\Hb) is mainly from the gravity of the central black hole.
The values of $r_{\rm{BLR}}/r_{\rm{g}}(f_{\rm{g}}=5.5)$ in Figure 3$a$ are estimated based on the uncorrected $M_{\rm{RM}}(f_{\rm{g}}=5.5)$. However, $M_{\rm{RM}}(f_{\rm{g}}=5.5)$ could not be corrected individually for each
quasar due to the absence of the different individual virial factor that is independent of $\Delta v$(\Hb).
The overall correction of $M_{\rm{RM}}(f_{\rm{g}}=5.5)$ for these 1973 quasars can be made by a factor of 3.4 derived
from $f_{\rm{g}}=5.5$ and an average of $f_{\sigma} =18.5$ presented in Figure 1$a$ (see Figure 3$b$). This overall correction is equivalent to the overall parallel shift of the data in Figure 3$a$. Figure 3$b$ shows that the negative correlation expectation is basically consistent with the trend between $\Delta v$(\Hb) and the corrected $r_{\rm{BLR}}/r_{\rm{g}}(f_{\rm{g}}=5.5)$. This indicates that $\Delta v$(\Hb) is dominated by the gravity of the central black hole. In addition, $r_{\rm{g}}/r_{\rm{BLR}}\lesssim 0.01 \ll 1$ for the $x$-axis values in Figures
3$a$ and 3$b$ and the values of $r_{\rm{g}}/r_{\rm{BLR}}$ corrected by $f_{\sigma}$ in Table 1. At these optical BLR
scales of quasars, the Schwarzschild metric is valid and still matches the weak field limit. The RM researches of AGNs
may shed light on the origin of $\Delta v$(\Hb), and the relevant discussion is presented in the next section.

The virial factor of \feii\ is larger than that of \Hb\ for 98\% of these quasars (see Figure 4). Also, Figure 4 shows
that the virial factor is very different from object to object and for different emission lines. If the stratified photoionization found for broad emission lines is prevalent in AGNs, the optimized photoionization zones in BLRs will be different for different lines. For clouds at a given radius, $F_{\rm{r}}$ is the resultant force from the different ions with negligible drift velocities between the gas constituents within these clouds, which generate a prominent broad emission line component. Also, the typical size of cloud of BLR is much less than the extent of BLR. So, $F_{\rm{r}}$ on clouds at the different radius is different, thus potentially resulting in different virial factors. \feii\ may be from a region outside the BLR of \Hb, i.e., $r_{\rm{BLR}}$(\feii) $ > r_{\rm{BLR}}$(\Hb) \citep[e.g.,][and references therein]{Hu08}. The RM of quasar 3C 273 showed $r_{\rm{BLR}}$(\feii) $> r_{\rm{BLR}}$(\Hb) \citep{Zh19}. For broad emission lines with different $r_{\rm{BLR}}$, there will be $f\propto r_{\rm{BLR}}^{\alpha}$ ($\alpha >0$) as $F_{\rm{r}}$ is considered and the BLR clouds are in the virialized motion for a given AGN \citep{Li17}. If $r_{\rm{BLR}}$(\feii) $> r_{\rm{BLR}}$(\Hb) and $f\propto r_{\rm{BLR}}^{\alpha}$, it will be expected that the measured virial factor of \feii\ is larger than that of \Hb\ for most of quasars in our sample.

\citet{Ne07} have suggested that $R_{\rm{FeII}}$ is a BLR metallicity indicator for SDSS type 1 AGNs. \citet{Pa18,Pa19}
have suggested that $R_{\rm{FeII}}$ is associated with the BLR metallicity. $R_{\rm{FeII}}$ increases with the increasing metallicity. The BLR metallicity can influence $F_{\rm{r}}$ due to the line-driven force dominated by the metallic elements \citep[e.g.,][]{Fe09,Da19}. Thus, $R_{\rm{FeII}}$ may influence the virial factor. Three positive correlations exist among $f_{\sigma}$, $R_{\rm{FeII}}$, and $\mathscr{\dot M}_{f_{\rm{g}}=5.5}$ (see Table 2 and Figure 5). Since three correlations exist among them, there should be a correlation like as $f_{\sigma}(\mathscr{\dot M}_{f_{\rm{g}}=5.5},R_{\rm{FeII}})$. In fact, there is a positive correlation at the confidence level of $>99.99\%$, $\log f_{\sigma}=-0.41+0.11\log\mathscr{\dot M}_{f_{\rm{g}}=5.5}+0.28\log R_{\rm{FeII}}$. Thus, $f_{\sigma}$ is dominated by $R_{\rm{FeII}}$ and $\mathscr{\dot M}_{f_{\rm{g}}=5.5}$ (or the Eddington ratio). This should be easily understood that $F_{\rm{r}}$ exerted on the BLR clouds will be larger as the BLR metallicity is higher and/or the radiation of accretion disk is stronger. Thus, the observed envelope delineating the data should be a consequence of physical effects, such as the Doppler effects, the gravitational redshift, and the line-driven force, which depend on the black hole mass, the bolometric luminosity of the black hole, and the BLR metallicity.

\begin{figure}
\begin{center}
\includegraphics[angle=-90,scale=0.28]{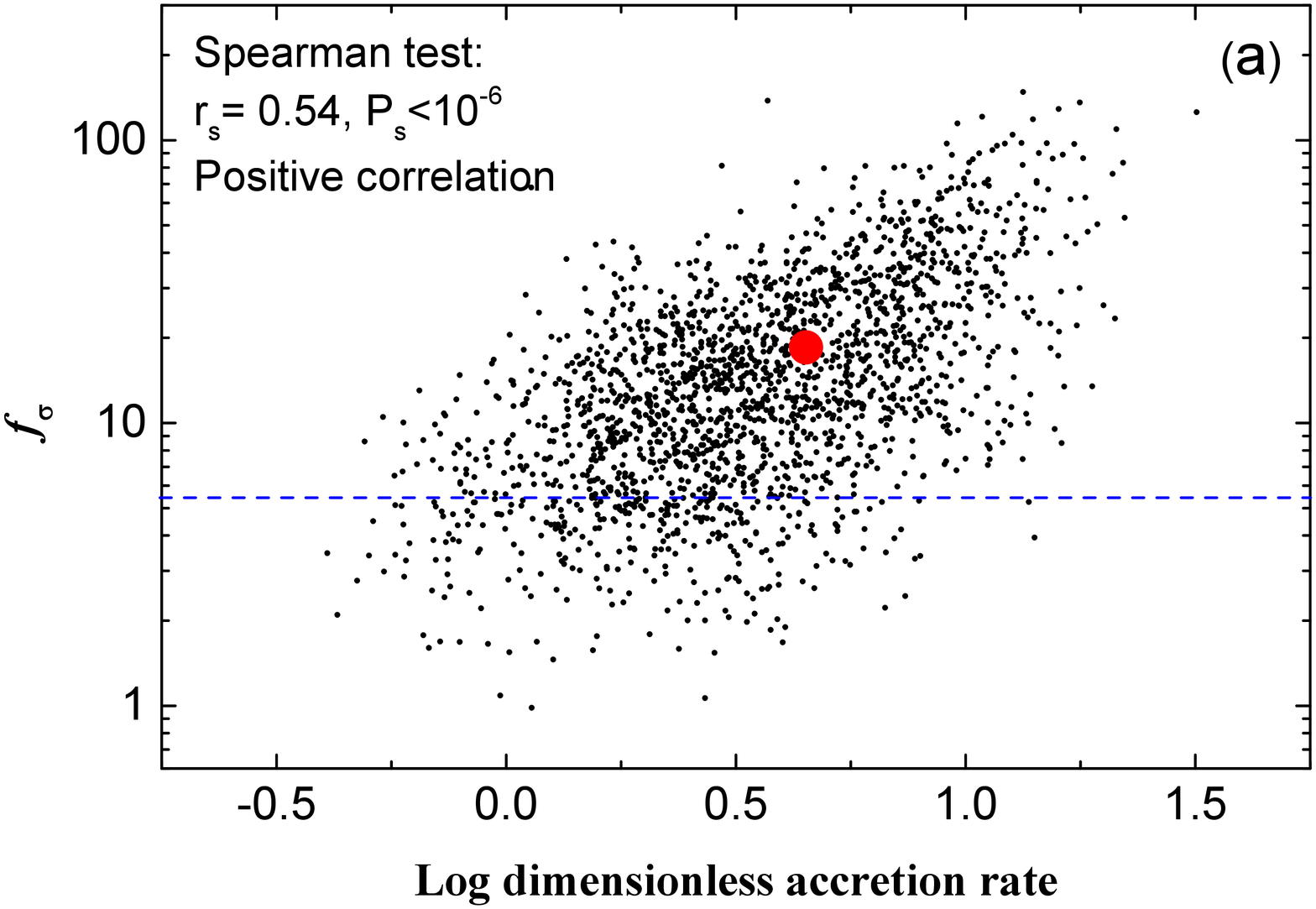}
\includegraphics[angle=-90,scale=0.28]{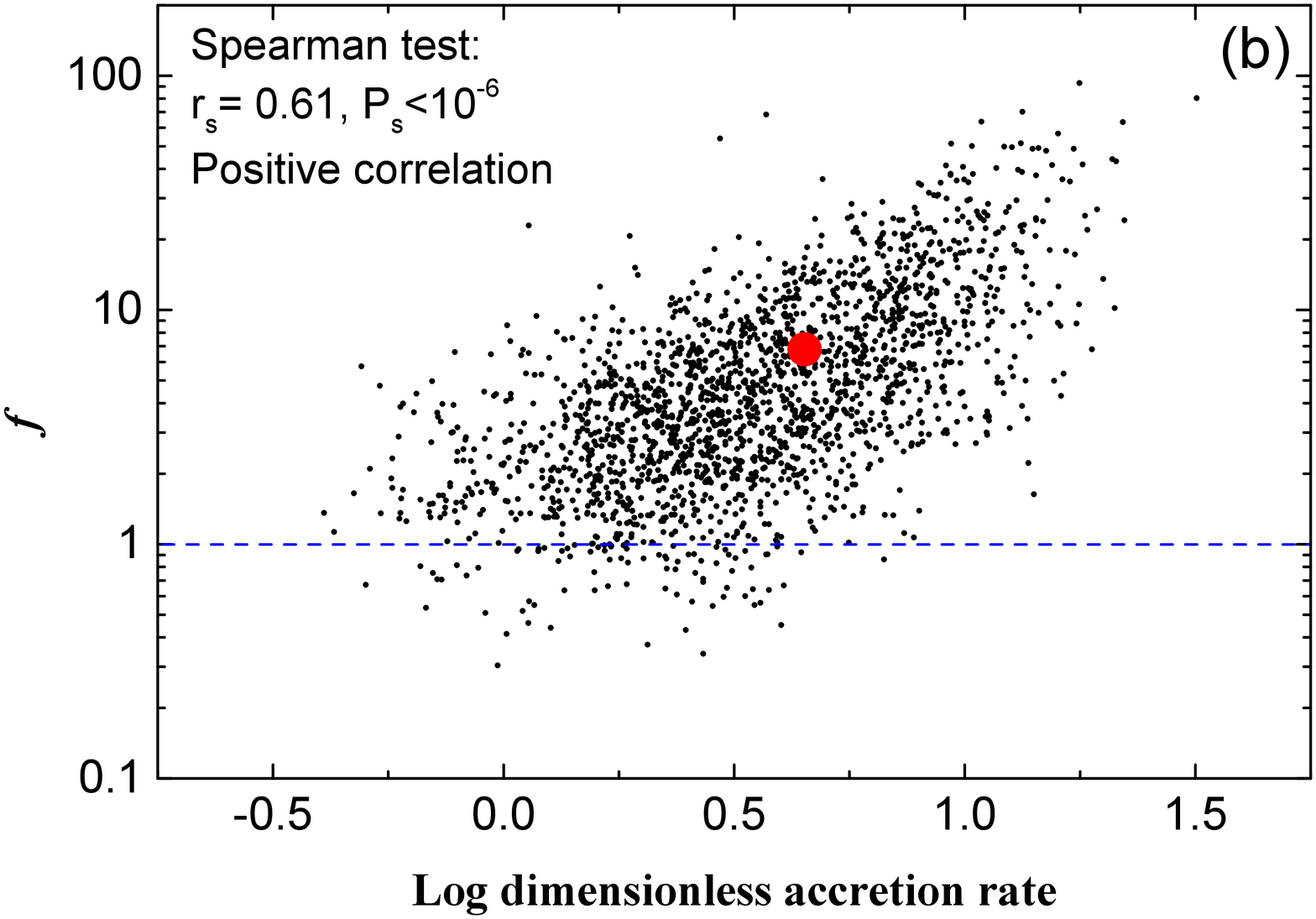}
\end{center}
\caption{Panel ($a$): \Hb-$\sigma_{\rm{line}}$-based $f_{\sigma}$ vs. $\mathscr{\dot M}_{f_{\rm{g}}=5.5}$. Spearman test shows a positive correlation between these two physical quantities, which have averages corresponding to the red solid circle. The dashed line denotes $f_{\rm{g}}=5.5$ for $\sigma_{\rm{line}}$. Panel ($b$): \Hb-$v_{\rm{FWHM}}$-based $f$ vs. $\mathscr{\dot M}_{f_{\rm{g}}=5.5}$. Spearman test shows a positive correlation between these two physical quantities, which have averages corresponding to the red solid circle. The dashed line denotes $f_{\rm{g}}=1$ for $v_{\rm{FWHM}}$.}
\label{fig1}
\end{figure}

\begin{figure}
\begin{center}
\includegraphics[angle=-90,scale=0.28]{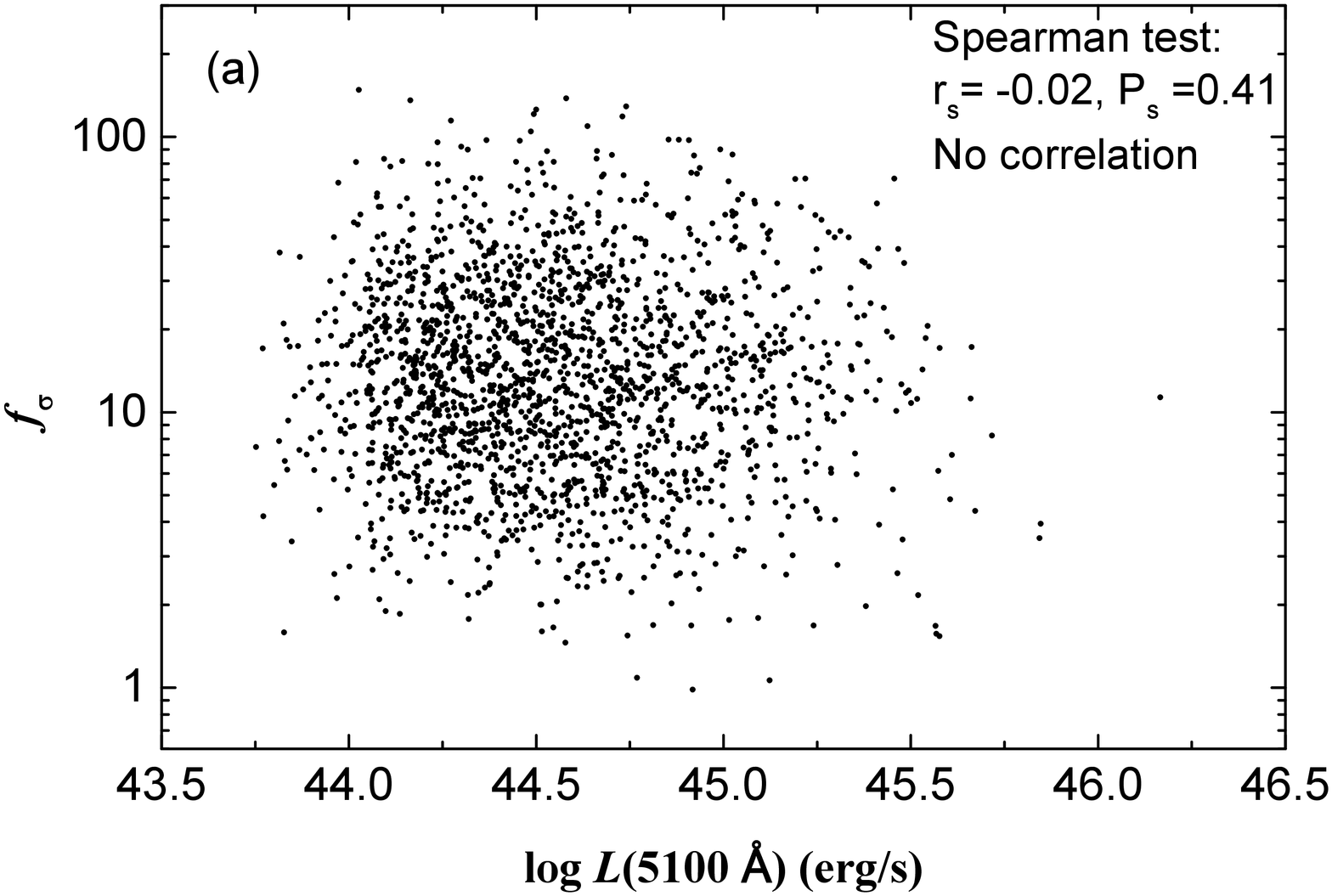}
\includegraphics[angle=-90,scale=0.28]{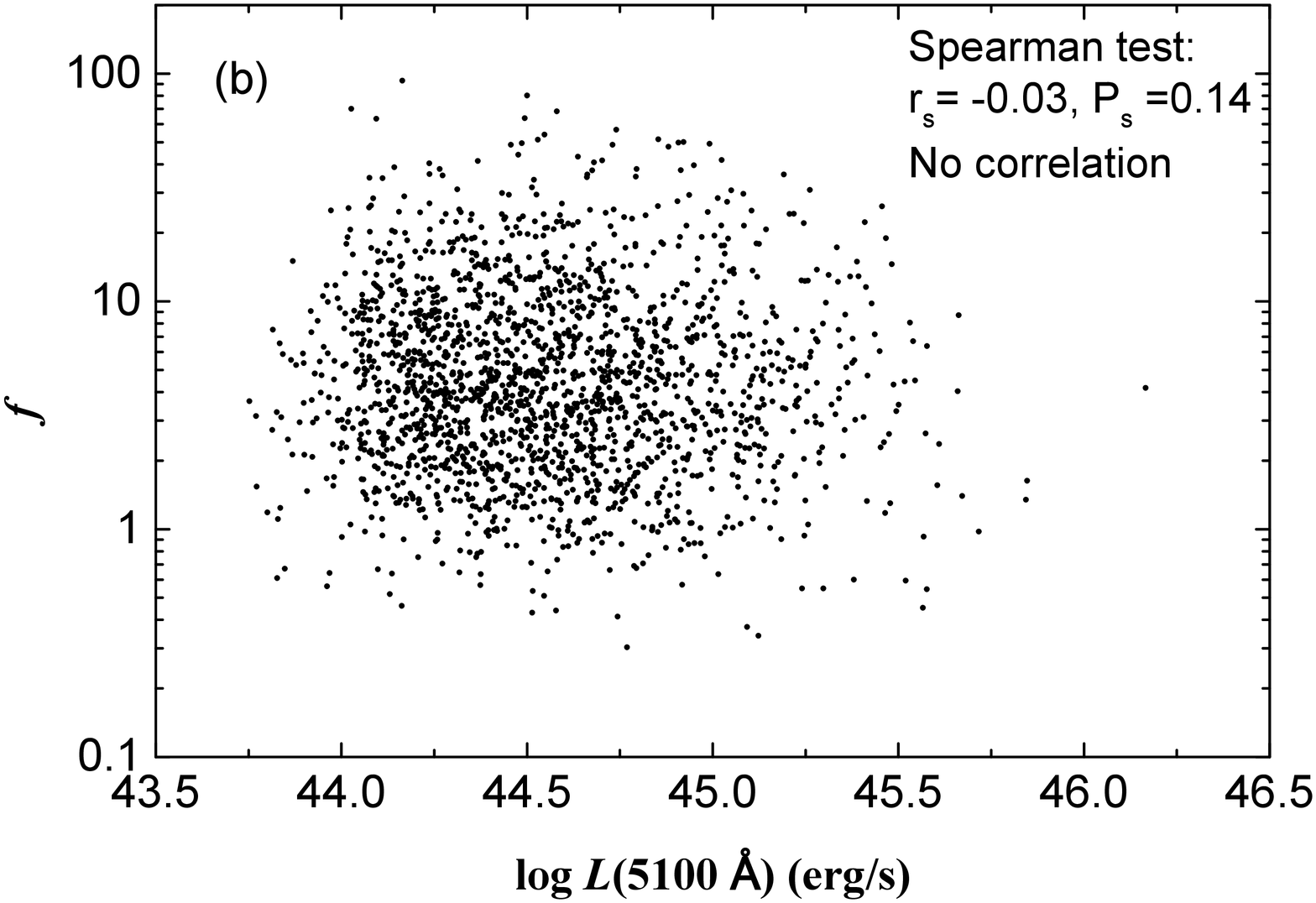}
\end{center}
\caption{Panel ($a$): \Hb-$\sigma_{\rm{line}}$-based $f_{\sigma}$ vs. $L(\rm{5100 \/\ \AA})$ and Panel ($b$): \Hb-$v_{\rm{FWHM}}$-based $f$ vs. $L(\rm{5100 \/\ \AA})$. Spearman test shows no correlation.}
\label{fig2}
\end{figure}

\begin{figure}
\begin{center}
\includegraphics[angle=-90,scale=0.28]{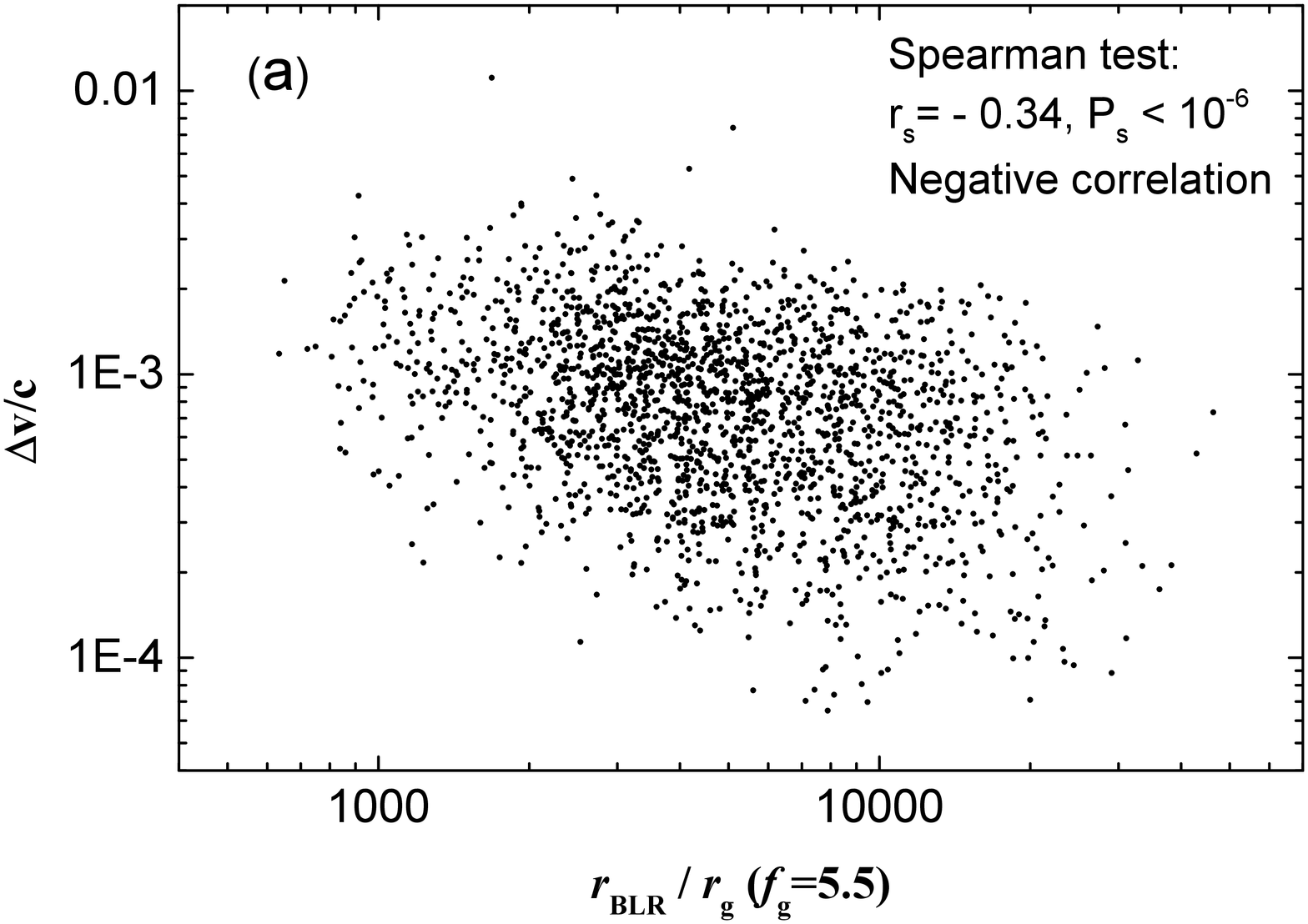}
\includegraphics[angle=-90,scale=0.28]{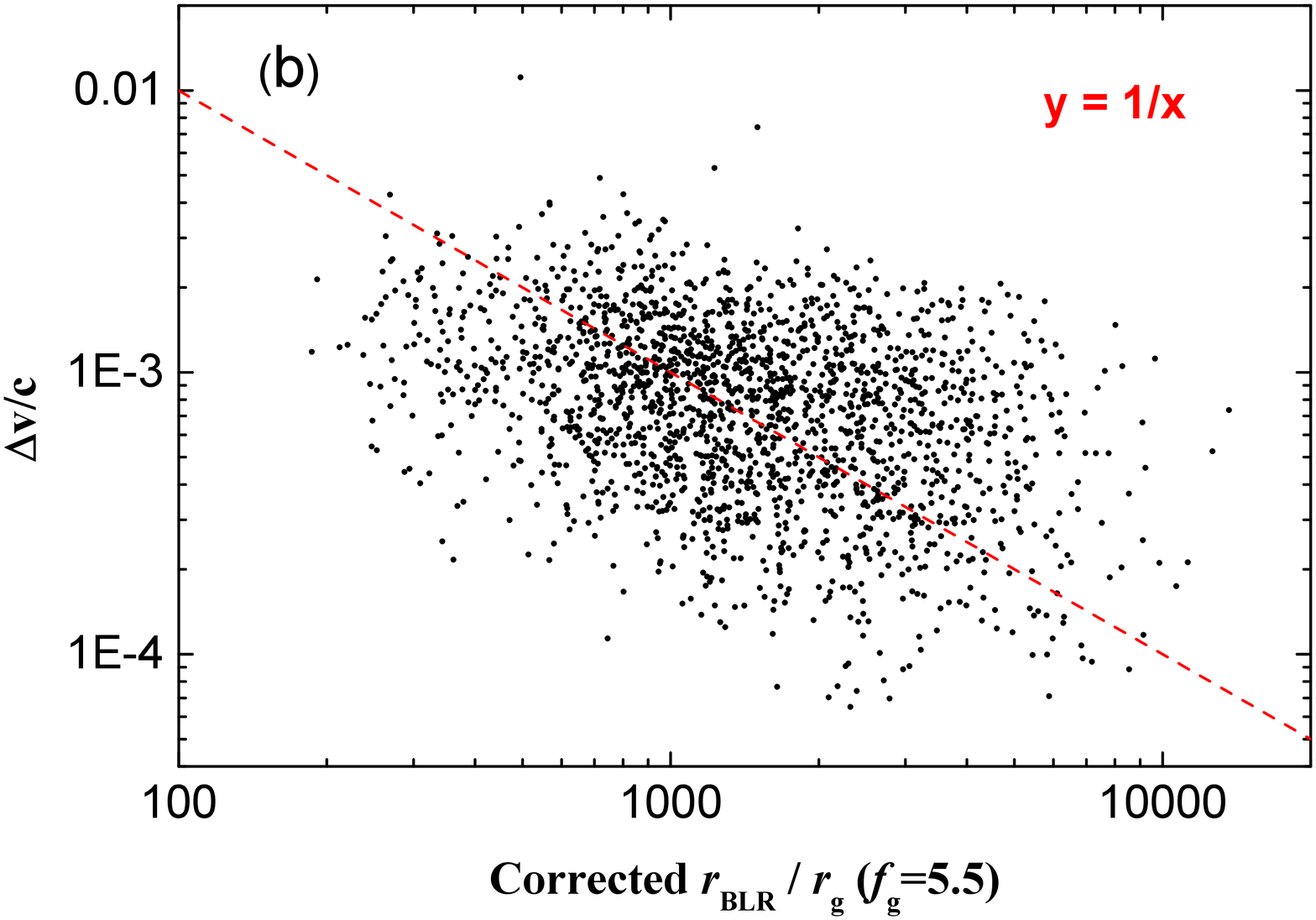}
\end{center}
\caption{Panel ($a$): \Hb\ velocity shift $\Delta v /c$ vs. $r_{\rm{BLR}}/r_{\rm{g}}(f_{\rm{g}}=5.5)$. Spearman test shows negative correlation between these two physical quantities. Panel ($b$): \Hb\ velocity shift $\Delta v /c$ vs. $r_{\rm{BLR}}/r_{\rm{g}}(f_{\rm{g}}=5.5)$ corrected by a factor of 3.4. }
\label{fig3}
\end{figure}

\begin{figure}
  \begin{center}
   \includegraphics[angle=-90,scale=0.40]{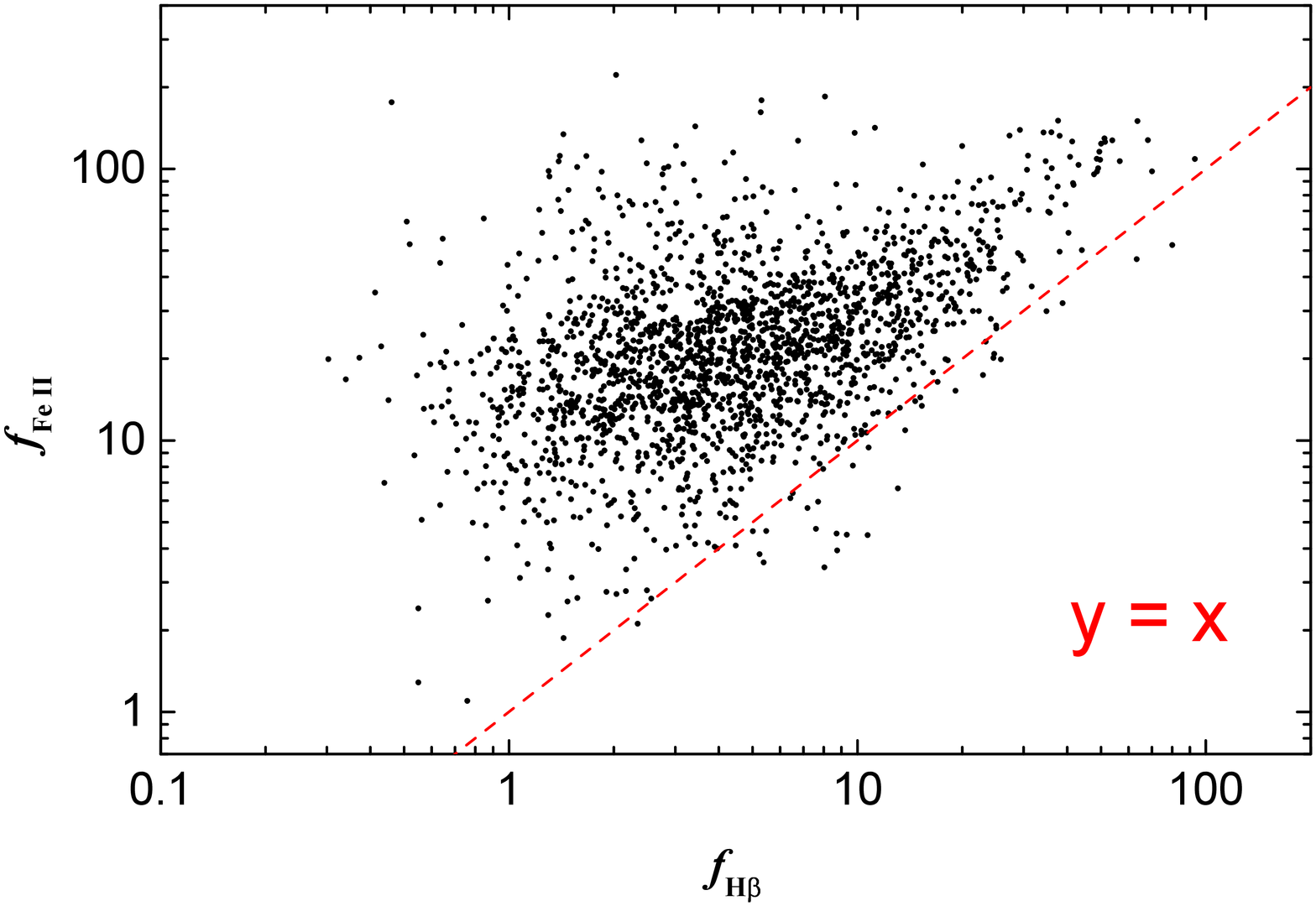}
  \end{center}
  \caption{$f$ for \feii\ vs. $f$ for \Hb\ based on $v_{\rm{FWHM}}$.}
  \label{fig4}
\end{figure}

\begin{figure}
  \begin{center}
   \includegraphics[angle=-90,scale=0.45]{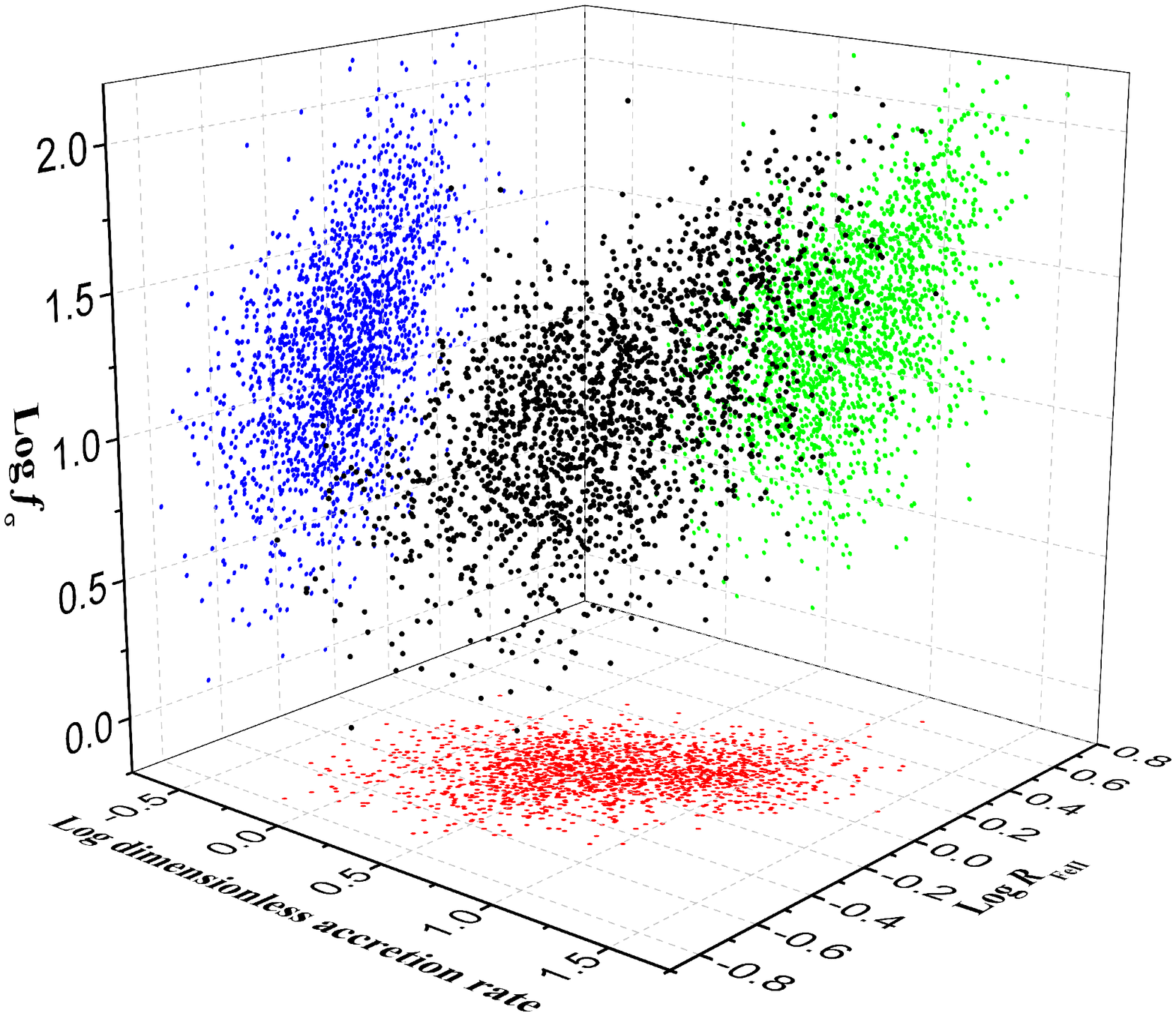}
  \end{center}
  \caption{3D plot of $\mathscr{\dot M}_{f_{\rm{g}}=5.5}$, $R_{\rm{FeII}}$, and $f_{\sigma}$ (black points). Color points correspond to $XY$, $XZ$, and $YZ$ projections of black points.}
  \label{fig5}
\end{figure}

\section{DISCUSSION AND CONCLUSIONS}

As $z_{\rm{g}} \ll 1$, Equation (4) can give for $\sigma_{\rm{line}}$, $f_{\sigma}$ and $z_{\rm{g}}=\Delta v/c$
\begin{equation}
  \log (\frac{\sigma_{\rm{line}}}{c})^2 = -\log(\frac{3}{2}f_{\sigma})+\log(\frac{\Delta v}{c}),
\end{equation}
which is similar to Equation (6) in \citet{MJ18}, where a tight correlation was found between the widths and redward shifts of the Fe {\sc iii}$\lambda\lambda$ 2039--2113 blend for their quasars, and this correlation supports the gravitational interpretation of the Fe {\sc iii}$\lambda\lambda$ 2039--2113 redward shifts. The Spearman's rank correlation test shows a positive correlation between the line width and velocity shift of the \Hb\ line for these 1973 quasars (see Table 2). A series of lines based on Equation (6) with different $f_{\sigma}$ are compared to the observational data points (see Figure 6). From top to bottom, the corresponding $f_{\sigma}$ increases. Because of the co-dependence between the Eddington ratio, dimensionless accretion rate and $\sigma_{\rm{line}}$, the large ranges of the former two quantities may lead to the large span in the direction roughly perpendicular to these lines (see Figure 6). Also, the metallicity difference of BLR might decrease correlations in Figure 1. Micro-turbulence within the BLR clouds can act as an apparent metallicity controller for the Fe {\sc ii}, and the reduction in the value of the metallicity can be up to a factor of ten for the Fe {\sc ii}/\Hb\ line ratio $R_{\rm{FeII}}$ when the micro-turbulence is invoked \citep{Pa21}. In addition, spectral simulations show that $R_{\rm{FeII}}$ depends clearly on cloud column density \citep[e.g.,][]{Fe09}. The combination of the column density, metallicity and internal physical processes may further decrease these correlations in Figure 1.
\begin{figure}
  \begin{center}
   \includegraphics[angle=-90,scale=0.40]{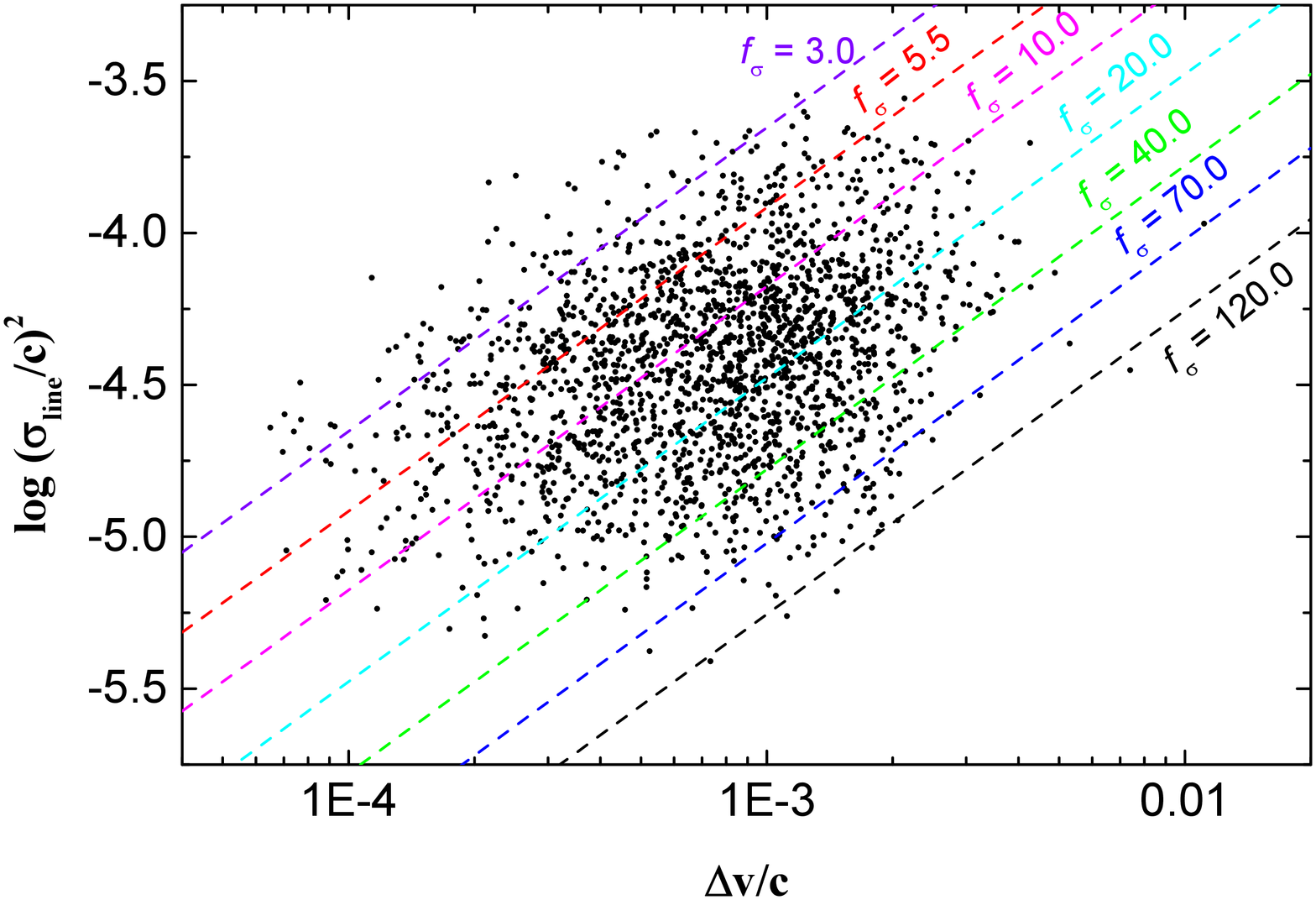}
  \end{center}
  \caption{$\log (\sigma_{\rm{line}}/c)^2$ vs. $\Delta v /c$ for the \Hb\ line. The values labelled on the dashed lines represent $f_{\sigma}$ in Equation (6).}
  \label{fig6}
\end{figure}

\begin{deluxetable}{cccc}
  \tablecolumns{4}
  \setlength{\tabcolsep}{5pt}
  \tablewidth{0pc}
  \tablecaption{Spearman's rank analysis results.}
  \tabletypesize{\scriptsize}
  \tablehead{\colhead{X}  &  \colhead{Y} &  \colhead{$r_{\rm{s}}$} &  \colhead{$P_{\rm{s}}$}}

  \startdata

 $\mathscr{\dot M}_{f_{\rm{g}}=5.5}$ or $L_{\rm{bol}}/L_{\rm{Edd}}$ & $f_{\sigma}$ & 0.54 &  $< 10^{-6}$   \\

 $\mathscr{\dot M}_{f_{\rm{g}}=5.5}$ or $L_{\rm{bol}}/L_{\rm{Edd}}$ & $f$  & 0.61  & $< 10^{-6}$      \\

 $L(\rm{5100 \AA})$ & $f_{\sigma}$ & -0.02  & 0.41   \\

$L(\rm{5100 \AA})$ &  $f$  & -0.03  & 0.14   \\

$r_{\rm{BLR}}/r_{\rm{g}}(f_{\rm{g}}=5.5)$ & $\Delta v /c$  & -0.34 &  $< 10^{-6}$   \\


 $\Delta v /c$     &  $  (\sigma_{\rm{line}}/c)^2$  & 0.34  &  $< 10^{-6}$  \\

 $R$(Fe {\sc ii}/\Hb)   &  $f_{\sigma}$   & 0.43  & $< 10^{-6}$  \\

 $\mathscr{\dot M}_{f_{\rm{g}}=5.5}$  or $L_{\rm{bol}}/L_{\rm{Edd}}$ &  $R$(Fe {\sc ii}/\Hb)   & 0.54  & $< 10^{-6}$   \\

\enddata
\tablecomments{\footnotesize X and Y are the relevant quantities presented in Figures 1--6. The SPEAR \citep{Pr92} gives $r_{\rm{s}}$ and $P_{\rm{s}}$ (the Spearman’s rank correlation coefficient and the p-value of hypothesis test, respectively).}
\label{Table2}
\end{deluxetable}

The shifts of broad emission lines may originate from the non-virialized BLR, e.g., outflows or inflows. However, gas outflows can generate the blueshifts of emission lines, e.g., narrow forbidden lines \oiii\ \citep[e.g.,][]{Co85,Sh14}
and broad emission line C {\sc iv} \citep[e.g.,][]{Wa11}. The \oiii\ emission lines can be decomposed into two components:
a narrow Gaussian component and a blueshifted/blue-skewed broad component. According to \citet{Hu08}, the shifts of broad emission lines are derived with respect to the narrow Gaussian component of \oiii$\lambda$ 5007. Thus, the shifts of broad emission lines taken from Table 2 in \citet{Hu08} are not influenced by these blueshifted/blue-skewed broad components of \oiii. The outflows in AGNs can be pushed by the line-driven force \citep[e.g.,][]{Da19,Dy18,Ma19},
and accretion disk winds driven by the line force have the increasing velocity with roughly decreasing acceleration
from the black hole to the far \citep{No20}. Observations show the various outflows at accretion disk scales, the BLR scales, the NLR scales and the kpc scales, driven by $F_{\rm{r}}$ from AGNs \citep{Ka18,Me21,Si21}. Thus, $F_{\rm{r}}$ is prevalent, and might contribute to the force budget for inflow, e.g., $F_{\rm{r}}$ decelerates inflow \citep{Fe09}. RM observations of PG 0026+129 indicate a decelerating inflow towards the black hole if $\Delta v$ originates from inflow. If the decelerating inflow is prevalent, $\Delta v$ will increase with the increasing $r_{\rm{BLR}}/r_{\rm{g}}(f_{\rm{g}}=5.5)$, and this expectation is not consistent with the trend found in Figure 3.
Thus, the inflow seems not to be the origin of $\Delta v$(\Hb).

It was believed that inflow generates the redward shifted broad emission lines with the blueward asymmetric velocity-resolved lag maps obtained in RM observations, and outflow generates the blueward shifted broad emission lines
with the redward asymmetric lag maps. However, the asymmetric lag maps and shifts of broad emission lines for AGNs usually differ from the expectations of inflow, such as 3C 273 \citep[e.g.,][]{Zh19}, PG 0026+129 \citep{Hu20}, NGC 3516 \citep[e.g.,][]{De10,Fe21a}, and NGC 2617 \citep[e.g.,][]{Fe21b}. The redward shifted broad emission lines with
the blueward asymmetric lag maps might be generated by an elliptical disklike BLR or a circular disklike BLR plus a spiral armlike BLR \citep{Fe21a}. Eccentricities and orientations of cloud orbits significantly influence full two-dimensional transfer function (2DTF) of a single disklike BLR \citep[see Figure 3 in][]{KW20}, and the redward shifted broad emission lines with various lag maps may originate from the clouds in virialized motion with various asymmetric responses in 2DTF. Virialized BLRs are suggested by the symmetric lag maps of redward shifted broad emission lines for SBS 1116+583A \citep{Be09}, Mrk 50 \citep{Ba11}, and SBS 1518+593 \citep{Du18a}. Therefore, the redward shifted broad emission lines in AGNs do not necessarily originate from inflow.

These 2485 quasars with $\Delta v$(\Hb)$>0$ in Table 2 of \citet{Hu08} have a median of $\Delta v$(\Hb)$ = 190
\rm{km \/\ s^{-1}}$ with a typical error of 55 $\rm{km \/\ s^{-1}}$, and an average and standard deviation of $\langle \Delta v \rangle$(\Hb)$=238 \pm 67$ $\rm{km \/\ s^{-1}}$. 1973 quasars in our sample have $\Delta v$(\Hb)$ = 242 \pm 52$ $\rm{km \/\ s^{-1}}$ and $\langle \Delta v \rangle$(\Hb)$=281 \pm 63$ $\rm{km \/\ s^{-1}}$. Considering uncertainties,
these distributions of $\Delta v$(\Hb) do not seem obviously different. The relevant distributions of $\Delta v$(\Hb)
are presented in Figure 7$a$. Sources with smaller $\Delta v$(\Hb) are more likely excluded by
$\Delta v - \sigma(\Delta v) > 0$, which means the relative error of $\Delta v$ is smaller than 1. The relative error distributions in Figure 7$b$ for the 1973 and 2485 quasar samples show that the relative error is more likely larger
for the smaller $\Delta v$. Correlation analyses for 2485 quasars show, at the confidence level of $>99.99\%$, three positive correlations among $f_{\sigma}$, $R_{\rm{FeII}}$ and $\mathscr{\dot M}_{f_{\rm{g}}=5.5}$, a positive correlation like as $f_{\sigma}(\mathscr{\dot M}_{f_{\rm{g}}=5.5},R_{\rm{FeII}})$, and a negative correlation between $\Delta v$(\Hb) and $r_{\rm{BLR}}/r_{\rm{g}}(f_{\rm{g}}=5.5)$. These results indicate that correlations found for 1973 quasars do not originate from the selection effect, i.e., $\Delta v - \sigma(\Delta v)> 0$ will not result in illusive correlations,
though this condition will make $\Delta v$ larger for the selected quasars. The fraction of blueshifted broad line \Hb\
with $\Delta v +3\sigma(\Delta v)<0$ is about 14\% for quasars in \citet{Hu08}, and these blueshifted quasars may be explained by additional blueshift of a kinematic origin arising from radial motion, e.g., outflows. Outflows seem exist
even if AGNs are during theirs low-flux states \citep[e.g.,][]{Me22}. Larger quasar sample, e.g., SDSS DR7 quasars in \citet{Li19} who gave the detailed parameters of spectra, will be used in the next work.
\begin{figure}
  \begin{center}
   \includegraphics[angle=-90,scale=0.28]{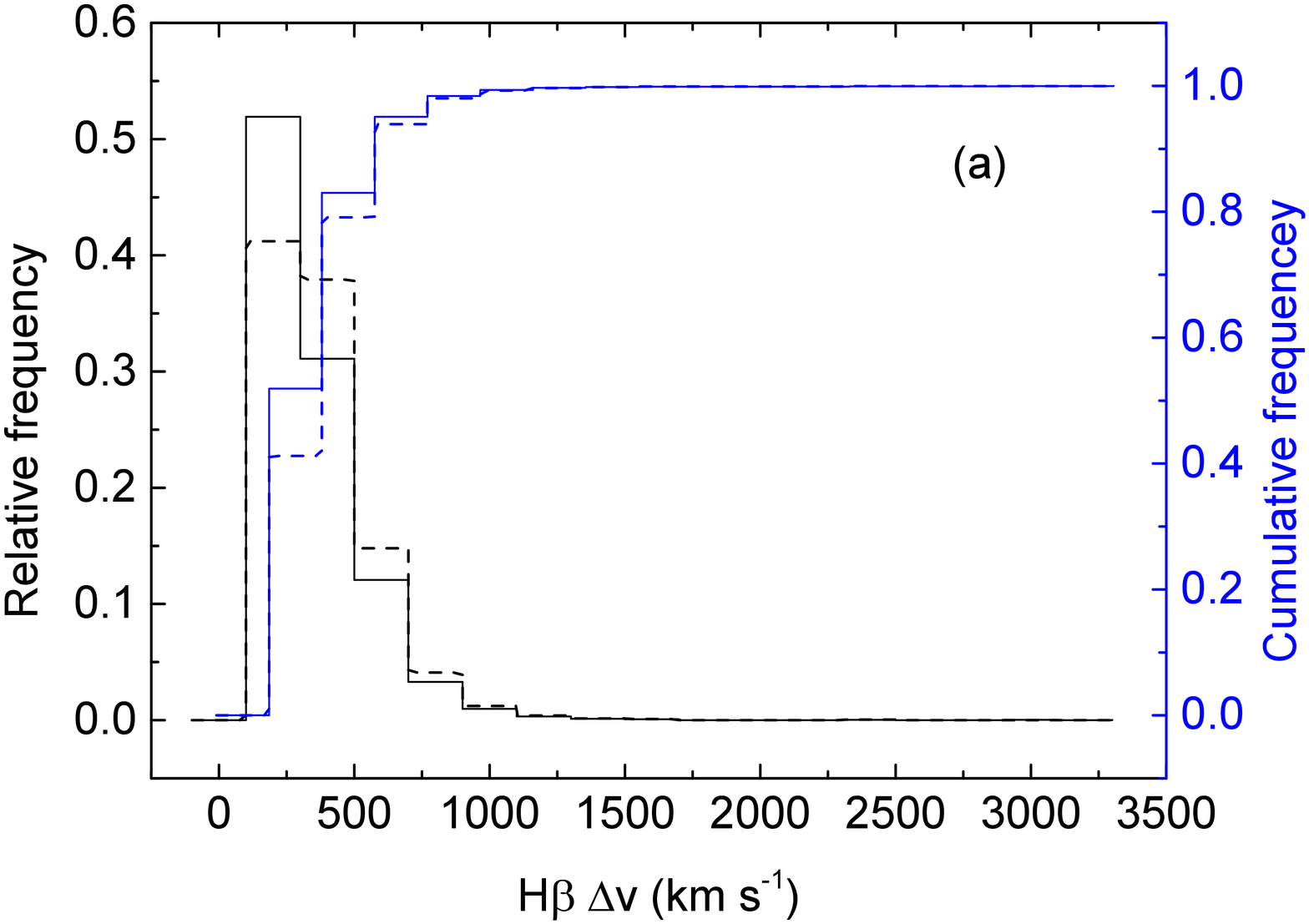}
   \includegraphics[angle=-90,scale=0.28]{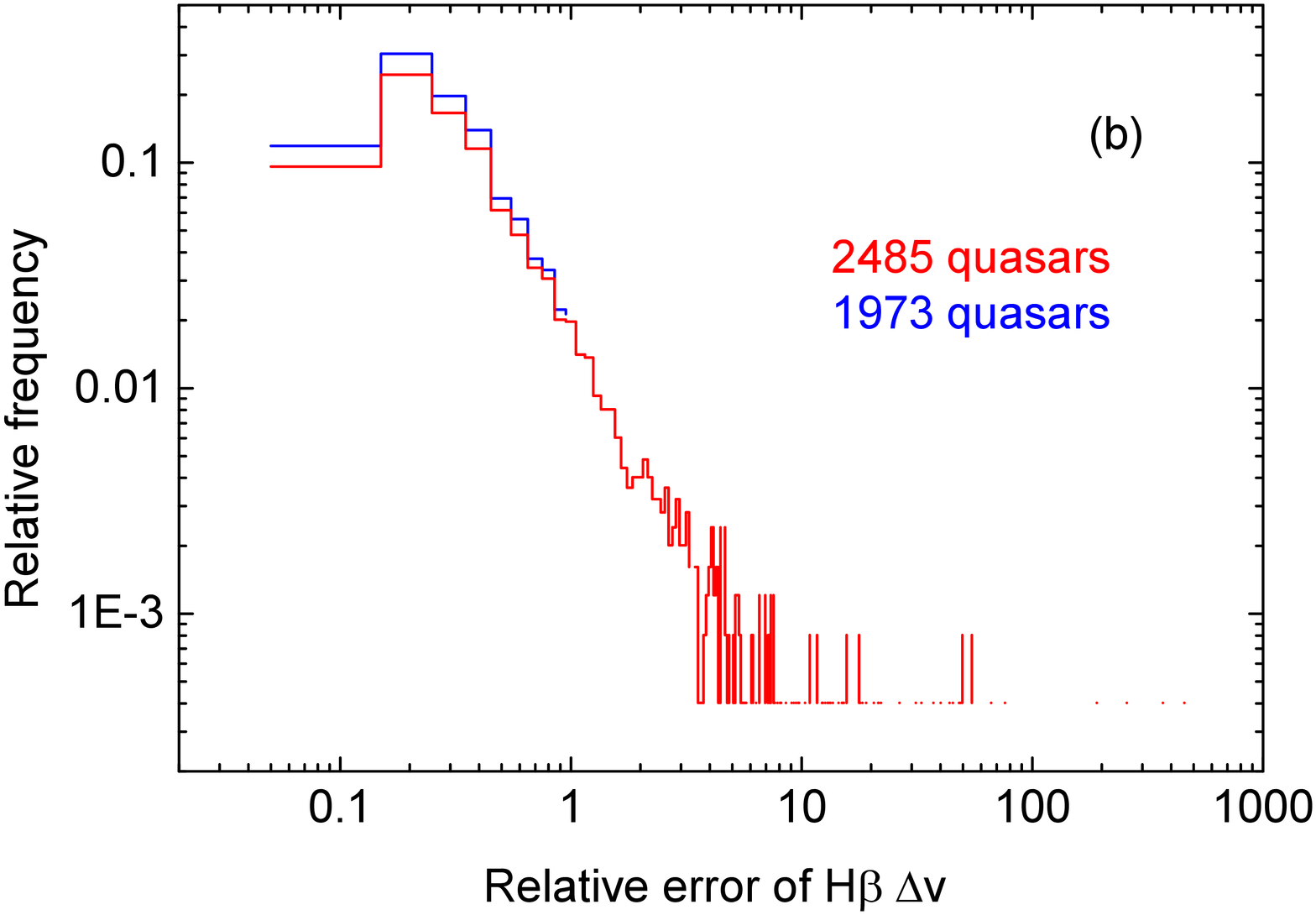}
  \end{center}
  \caption{Distributions of $\Delta v$(\Hb) and its relative error. Panel ($a$): the solid lines are the distributions for 2485 quasars with $\Delta v$(\Hb)$>0$ in \citet{Hu08}. The dashed lines are the distributions for 1973 quasars in our sample. Panel ($b$): distributions of the relative error of $\Delta v$(\Hb).}
  \label{fig7}
\end{figure}

The radiative efficiency is closely related to a black hole spin, but it is difficult to measure the spin of the black hole in AGN. Usually, the Eddington ratio is regarded as a proxy of accretion rate of black hole. Even though these correlations of the dimensionless accretion rate with the other physical quantities are likely influenced by the unknown real individual value of radiative efficiency, there are still correlations of the Eddington ratio with these physical quantities, because only a difference of 0.038 exists between $\mathscr{\dot M}_{f_{\rm{g}}=5.5}$ and $L_{\rm{bol}}/L_{\rm{Edd}}$ in Table 1. Based on $L_{\rm{bol}}$ and the mass accretion rate, \citet{Da11} determined $\eta$ for a sample of 80 Palomar–-Green quasars, and found a strong correlation of $\eta=0.089 M_{8}^{0.52}$, where $M_{8}$ is the black hole mass in units of
$10^8 M_{\odot}$. In order to test the influence of a fixed radiative efficiency $\eta=0.038$, this empirical relation is used to estimate $\eta$. Correlation analyses are made for those quantities in Figures 1 and 5 with $\mathscr{\dot M}_{f_{\rm{g}}=5.5}$ to be re-estimated by $L_{\rm{bol}}/L_{\rm{Edd}}$ in Table 1 and the estimated $\eta$. There are still correlations very similar to those found in Figures 1 and 5 as using these new dimensionless accretion rates (see Figures 8 and 9). A positive correlation, $\log f_{\sigma}=-0.24+0.09\log\mathscr{\dot M}_{f_{\rm{g}}=5.5}+0.21\log R_{\rm{FeII}}$, exists at the confidence level of $>99.99\%$ (see Figure 9). Thus, these correlations found in this work do not result from using the fixed value of $\eta=0.038$.
\begin{figure}
  \begin{center}
   \includegraphics[angle=-90,scale=0.28]{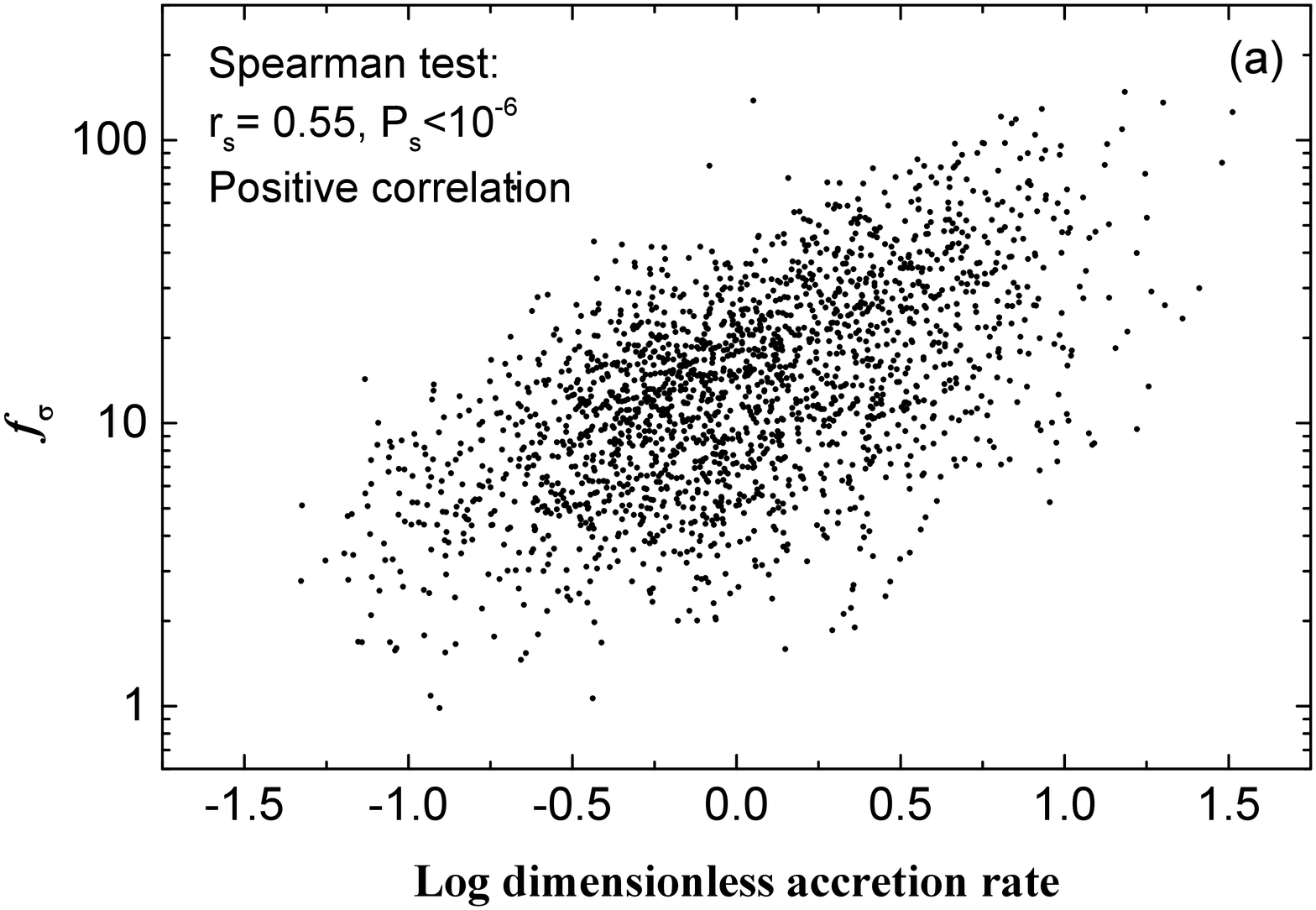}
   \includegraphics[angle=-90,scale=0.28]{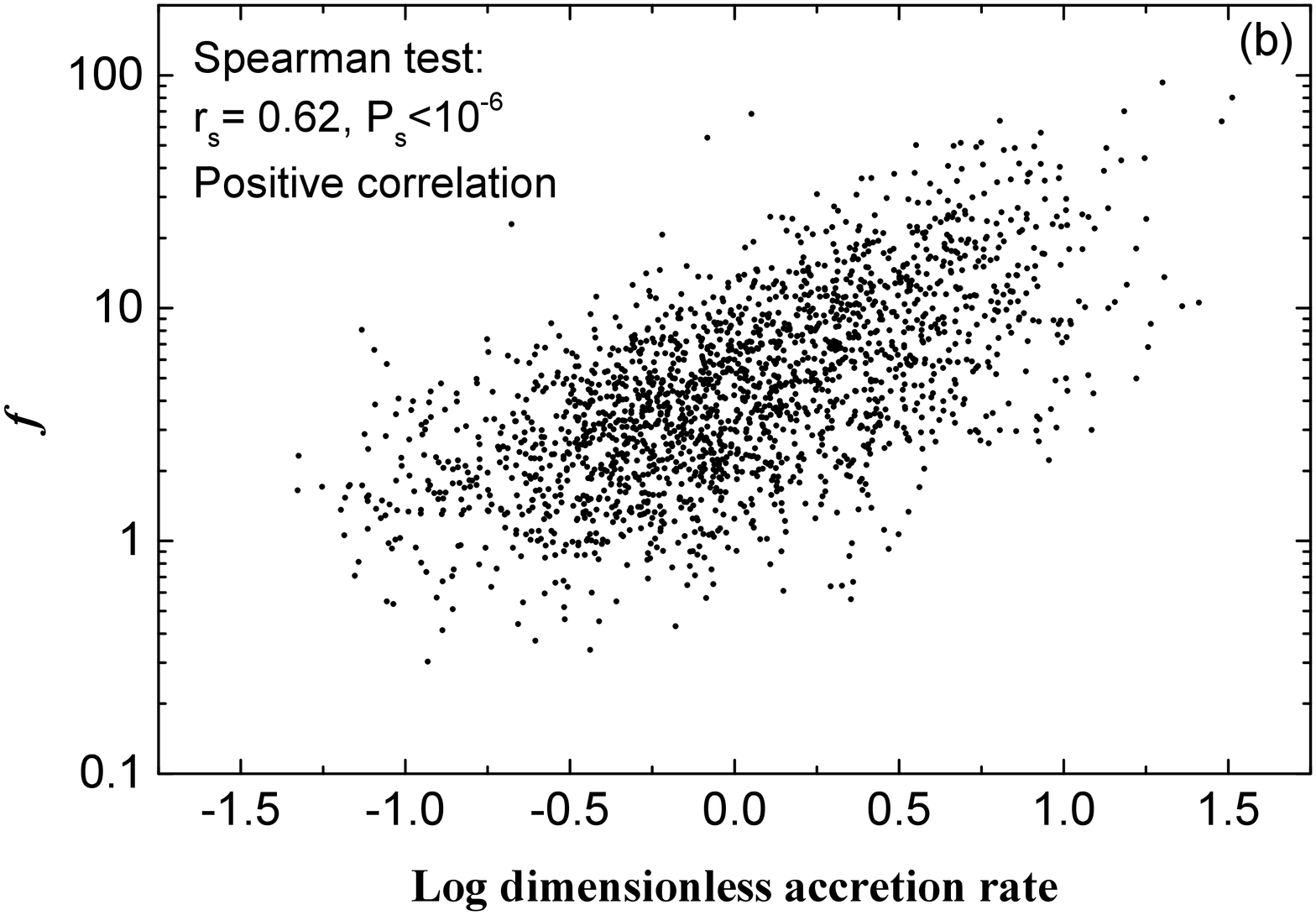}
  \end{center}
  \caption{Panel ($a$): \Hb-$\sigma_{\rm{line}}$-based $f_{\sigma}$ vs. $\mathscr{\dot M}_{f_{\rm{g}}=5.5}$. Spearman test shows a positive correlation between these two physical quantities. Panel ($b$): \Hb-$v_{\rm{FWHM}}$-based $f$ vs. $\mathscr{\dot M}_{f_{\rm{g}}=5.5}$. Spearman test shows a positive correlation between these two physical quantities. $\mathscr{\dot M}_{f_{\rm{g}}=5.5}$ is the value estimated by $\eta=0.089M_{8}^{0.52}$ rather than $\eta=0.038$.}
  \label{fig8}
\end{figure}

\begin{figure}
  \begin{center}

   \includegraphics[angle=-90,scale=0.4]{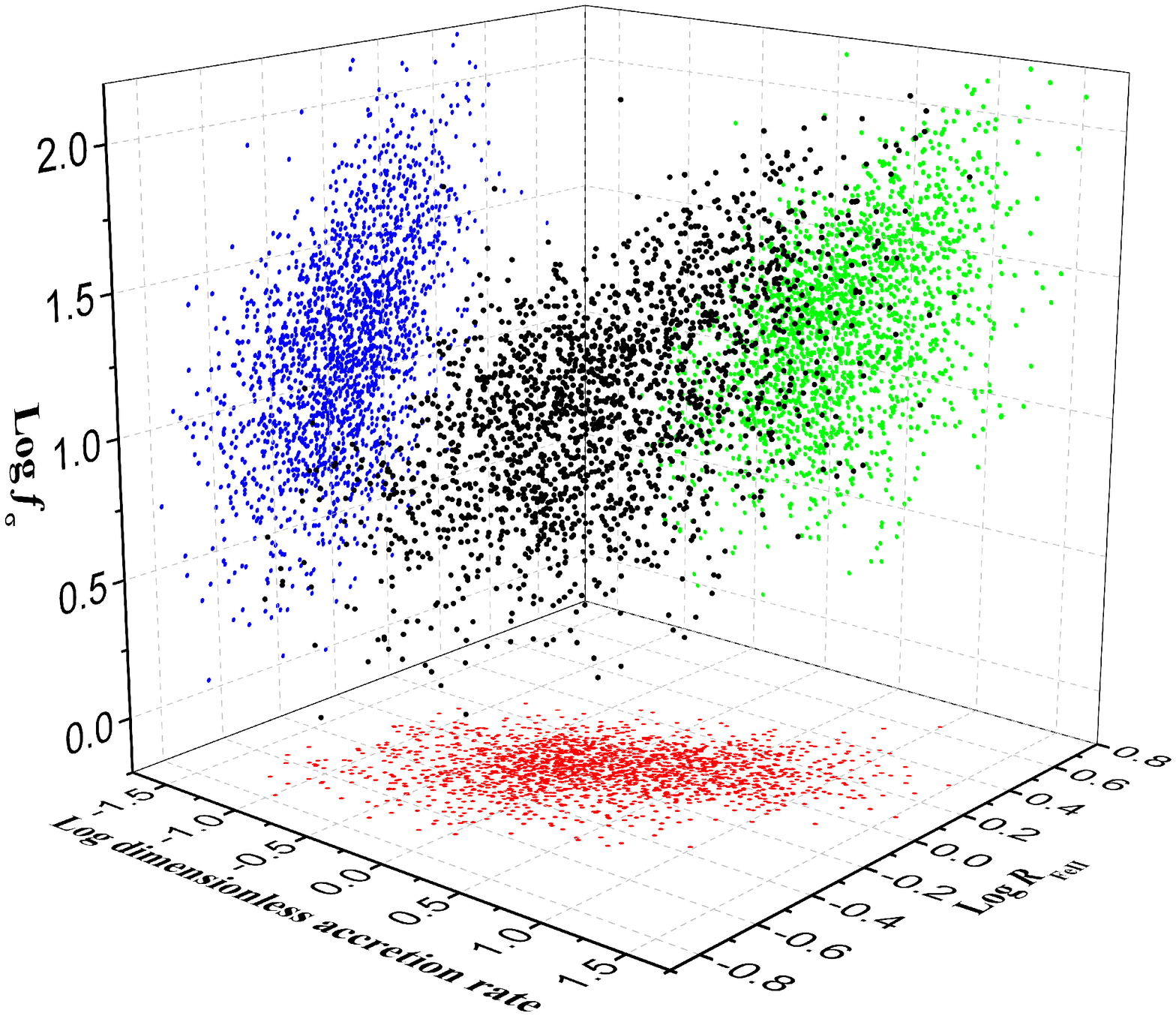}
  \end{center}
  \caption{3D plot of $\mathscr{\dot M}_{f_{\rm{g}}=5.5}$, $R_{\rm{FeII}}$, and $f_{\sigma}$ (black points), where $\mathscr{\dot M}_{f_{\rm{g}}=5.5}$ is the value estimated by $\eta=0.089M_{8}^{0.52}$ rather than $\eta=0.038$. Color points correspond to $XY$, $XZ$, and $YZ$ projections of black points.}
  \label{fig8}
\end{figure}

Based on the assumption of a gravitational origin for the redward shifts of broad emission lines \Hb\ and \feii, and their widths and redward shifts for a sample of 1973 $z<0.8$ SDSS DR5 quasars, we measured the virial factor in $M_{\rm{RM}}$, estimated by the RM method and/or the relevant secondary methods. The measured virial factor contains the overall effect
of $F_{\rm{r}}$ from accretion disk radiation and the geometric effect of BLR. $f_{\rm{\sigma}}>5.5$ and $f>1$
for the broad \Hb\ in most quasars (see Figure 1). $f_{\rm{FeII}}>f_{\rm{H\beta}}$ for 98\% of these 1973 quasars, which
is consistent with the deduction from $r_{\rm{BLR}}$(\feii) $> r_{\rm{BLR}}$(\Hb) and $f\propto r_{\rm{BLR}}^{\alpha}$ ($\alpha >0$). The virial factor is very different from object to object and for different emission lines (see Figure 4).
A series of lines, based on Equation (6) with different $f_{\sigma}$, basically reproduce the distribution of $(\sigma_{\rm{line}},\Delta v)$ for the \Hb\ line (see Figure 6), supporting the gravitational interpretation of $\Delta v$ for the \Hb\ line. There are three positive correlations among $f_{\sigma}$(\Hb), $\mathscr{\dot M}_{f_{\rm{g}}=5.5}$ and $R_{\rm{FeII}}$. A correlation, $\log f_{\sigma}=-0.41+0.11\log\mathscr{\dot M}_{f_{\rm{g}}=5.5}+0.28\log R_{\rm{FeII}}$, indicates that the virial factor is dominated by $\mathscr{\dot M}_{f_{\rm{g}}=5.5}$ and metallicity, which will influence $F_{\rm{r}}$ on the BLR clouds. $\Delta v$(\Hb) is anti-correlated with $r_{\rm{BLR}}/r_{\rm{g}}(f_{\rm{g}}=5.5)$, supporting the gravitational origin of the redward shift of \Hb. Our results indicate that $F_{\rm{r}}$ may be an important contributor to the virial factor, and the redward shifted broad emission lines show the potential of measuring the virial factor.

\acknowledgements {We are very grateful to the anonymous referees for constructive comments leading to significant improvement of this paper. We thank the helpful discussions of Prof. J R Mao. We thank the financial support of
the National Natural Science Foundation of China (NSFC; grant No. 11991051), and we acknowledge the science research
grants from the China Manned Space Project with NO. CMS-CSST-2021-A06.}


\begin{center}
\textbf{ORCID iDs}
\end{center}
H. T. Liu https://orcid.org/0000-0002-2153-3688 \\
Hai-Cheng Feng https://orcid.org/0000-0002-1530-2680

\label{lastpage}

\end{document}